\documentclass[10pt,conference]{IEEEtran}

\usepackage{color}

\usepackage[latin1]{inputenc}

\usepackage[pdftex]{graphicx}

\IEEEoverridecommandlockouts

\begin{document}

\title{NCSA: A New Protocol for Random Multiple Access Based on Physical Layer Network Coding}

\author{
\authorblockN{Huyen-Chi Bui$^{1,2}$, Jérôme Lacan$^1$ and Marie-Laure Boucheret$^2$}
\authorblockA{$^1$University of Toulouse, ISAE/DMIA, \\
$^2$University of Toulouse, IRIT/ENSEEIHT\\
Email: \{huyen-chi.bui,jerome.lacan\}@isae.fr, marie-laure.boucheret@n7.fr
}
}

\maketitle

\begin{abstract}
This paper introduces a random multiple access method for satellite communications, named Network Coding-based Slotted Aloha (NCSA). The goal is to improve diversity of data bursts on a slotted-ALOHA-like channel thanks to error correcting codes and Physical-layer Network Coding (PNC). This scheme can be considered as a generalization of the Contention Resolution Diversity Slotted Aloha (CRDSA) where the different replicas of this system are replaced by the different parts of a single word of an error correcting code.  The performance of this scheme is first studied through a density evolution approach. Then, simulations confirm the CRDSA results by showing that, for a time frame of $400$ slots, the achievable total throughput is greater than $0.7\times C$, where $C$ is the maximal throughput achieved by a centralized scheme. This paper is a first analysis of the proposed scheme which open several perspectives. The most promising approach is to integrate collided bursts into the decoding process in order to improve the obtained performance.
\end{abstract}
\section{Introduction and Related Work}
\label{introduction}
The success of random multiple access protocols, e.g. ALOHA \cite{abramson:aloha} and its variants, have motivated an extremely large number of studies. The pure Aloha is a protocol developed at the University of Hawaii for sharing channel access among a number of users with relatively low throughput demand. In such system, each user wishing to use a channel sends at any time short data packets independently from the others. When at least two users emit a packet simultaneously, a collision occurs. These packets are considered as lost and must be re-transmitted later. The optimal throughput is equal to $0.18 \times C$, where $C$ denotes the throughput of a centralized scheme. Today, different versions of Aloha (e.g. Slotted Aloha (SA) \cite{paperSA} or its enhanced version named Diversity Slotted Aloha (DSA) \cite{paperDSA}) are used in satellite networks for transmission of short bursts. In SA, the users send packets at fixed time slots of one packet length, and the maximum normalized throughput is doubled compared to the pure Aloha protocol (0.37 vs. 0.18). Utilization of SA is however limited to the transmission of signaling packets due to the large resulting propagation delay. The DSA protocol is an improved version of SA, used in TDMA or Multi-Frequency TDMA systems. In DSA, the same packet is transmitted twice to improve throughput performance and delay. However, the difference in performance of these protocols is pretty poor. In the subsequent evolution, recently, enhanced versions of DSA named Contention Resolution Diversity Slotted Aloha (CRDSA) \cite{paperCRDSA} and CRDSA++ \cite{paperCRDSA++} have been developed by the European Space Agency. In the CRDSA protocol, two replicas (3 to 5 replicas in the case of CRDSA++) of the same packet are generated and sent randomly onto the time frame. The improvement is that each packet contains a signaling information which points to its replicas location. When one packet is successfully decoded, the replicas are also located and cancelled. The CRDSA and CRDSA++ (called subsequently CRDSA*) decoding processes use Successive Interference Cancellation (SIC). CRDSA* appear to be efficient methods for satellite networks, especially for a Digital Video Broadcasting Return Channel via Satellite (DVB-RCS) scheme. The objective of our paper is to generalize the underlying idea proposed in CRDSA* and to formalize the results. 

Concerning the subtraction of signal in decoding algorithms, recent practical systems have shown that, under certain conditions, collided data can be exploited when one of the data is known. The Viasat's technology, called Paired Carrier Multiple Access (PCMA) \cite{dankberg:pcma}, allows two different earth stations to use simultaneously the same frequency, time slots, and/or CDMA code. This ViaSat-exclusive technique can increase satellite bandwidth capacity by as much as 100\%. A comparable approach, named Physical-Layer Network Coding (PNC) for wireless networks \cite{zhangLL06:plnc} \cite{analogNC} and recently in satellite context \cite{paperDLR} has also been proposed.

Amongst related work, several studies have shown that interference cancellation techniques bring out benefits in the context of random multiple access. As an example, in \cite{access:Cui} the authors study successive interference cancellations applied to each time slot. In network coding area, the joint use of classical network coding (i.e. performing operations on bits and not on signals) and random multiple access methods has been proposed and analyzed in \cite{NC2hops:aloha}. 
\begin{figure}
\centering
\includegraphics[width=4cm]{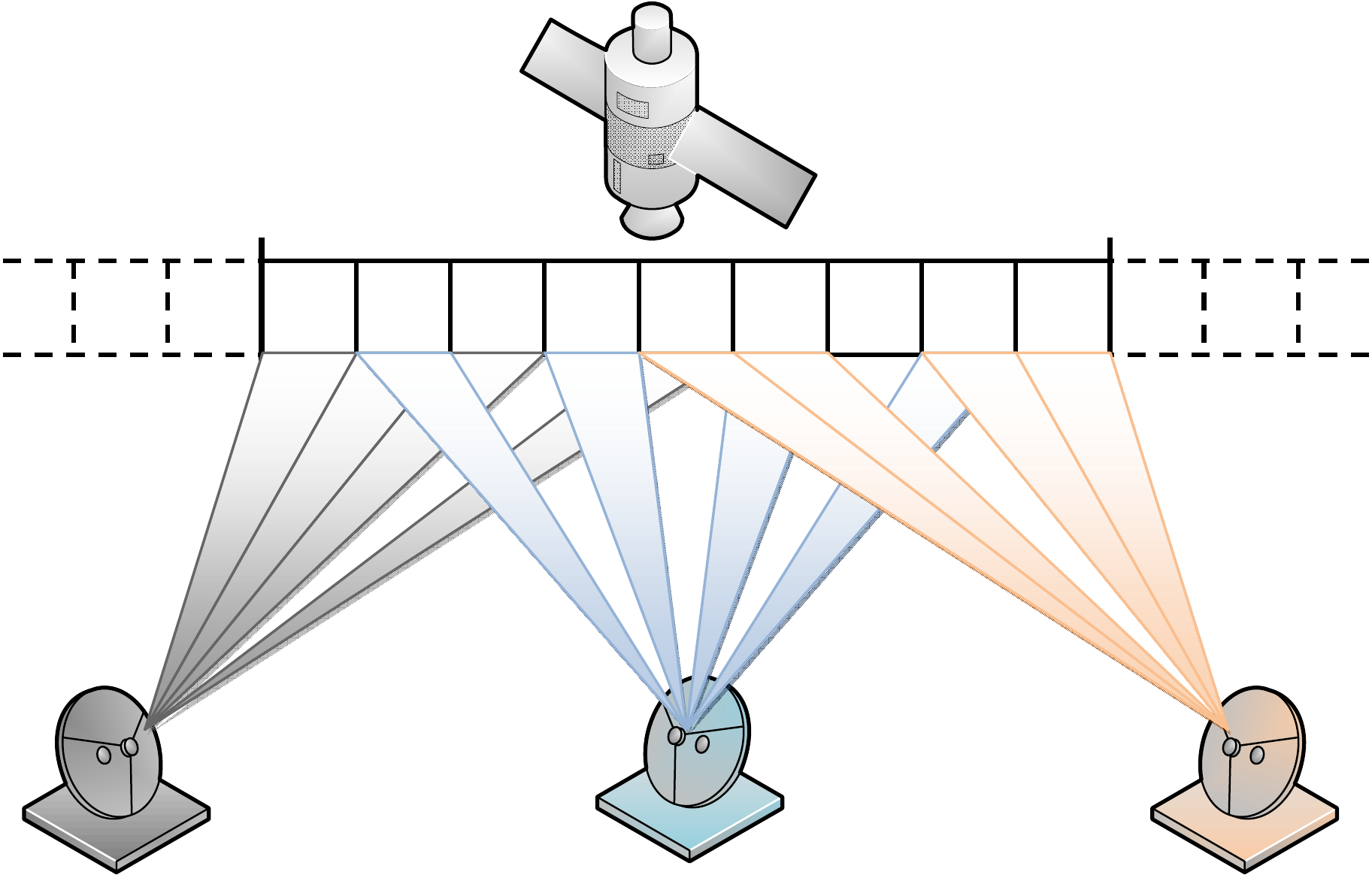}
\caption{Multiple access on a slotted channel}
\label{fig:system}
\end{figure}
Note that the NCSA method is not restrained to the domain of satellite communications. In this paper, we consider a system where several users share a channel to send data to a given satellite (see Figure \ref{fig:system}). All users are assumed to generate one data block, each one is cut into $n$ physical layer packets of same length called bursts. Like slotted-ALOHA, the frame is split into time slots of one burst duration. Then, the user randomly chooses $n$ slots and sends $n$ bursts on these slots. The receiver (which can be either embedded in the satellite, in the gateway or in the terminal) observes collisions on some slots. To recover collided data, it considers independently the data of each terminal. Similarly to CRDSA*, bursts collisions are almost cleaned up by the successive interference cancellation operations.


We have modeled the decoding process by using density evolution methods, classically applied in the context of LDPC decoding \cite{ldpccode:introduction}. Indeed, the data recovery process can be considered as a message-passing algorithm where some messages are exchanged between the user nodes and the slot nodes. The theoretical results are validated by simulations. We show (under our hypothesis) that this system can reach a throughput greater than $0.6\times C$.

The paper is organized as follows: the proposed multiple access scheme is detailed and compared to the CRDSA* protocols in the following section. In section~\ref{sec:evaluation}, this mechanism is evaluated by a density evolution approach. Then, simulation results are presented to validate the theoretical analysis and to estimate the maximum throughput of the system. Future work and conclusion are given in the last sections.
\section{Description of the Scheme}
\label{sec:description}
In this section we first present the assumptions taken on the communication system. Then, we detail the decoding mechanism used on the receiver which combines classical error decoding and SIC.
\subsection{Hypotheses}
\label{sec:hypotheses}
We consider that the wireless network system includes a number of users who communicate through a satellite (see Figure \ref{fig:system}). Therefore, the transmission is subject to an Additive White Gaussian Noise (AWGN). The access method to the satellite is based on slotted Aloha where time is divided into slots of one burst duration. The burst size is the same for all users. We consider that a set of $N_s$ consecutive time slots forms a time frame.

In our system, a user can only send $n$ bursts on a time frame. To continue to send more messages, the user must wait until the beginning of the next frame. The generalization is straight forward. We assume that synchronization mechanisms allow the users to be synchronized at slots and frames time levels.

Upon the transmission to the satellite, bursts of all users are mapped onto slots. If several nodes attempt a transmission during a slot, there is a collision. 
In this work, collided bursts are considered as erased. We assume that, with the used coding rate of the error correcting code, if $k$ bursts amongst $n$ of one user are uncollided, the resulting message of $n$ bursts is decodable by the receiver.
\subsection{Principle of the Mechanism}
\label{sec:principle}
In this system, each user wishes to transmit a data block of $L$ bits on each frame. Before sending data, it first uses an error-correcting code of rate $R$ able to cope with the erasures caused by the collisions and the errors caused by the noise on uncollided bursts. An good example of such code is the 3GPP2 turbo code, also used in DVB-SH, which manages simultaneously errors and erasures \cite{DVBSH-guidelines}. A block of $R\times L$ bits is generated. Encoded data block is then interleaved and split into $n$ bursts. Similarly to CRDSA*, the same preamble and signaling information bits are added to each burst. Modulated bursts are sent on several slots of a slotted-ALOHA-like channel. The signaling bits contain information identifying the position of other bursts of same user within the time frame. The preamble is a pseudo-random sequence and unique to each user. Given a target $E_b/N_0$ and a set of parameters $(n,~k)$, the code rate $R$ is determined in order to guarantee a given Bit Error Rate (BER) when $n-k$ bursts among $n$ are erased (\textit{i.e.} collided). We assume that all users have the same code which is known by the receiver. In the CRDSA* scheme, a terminal sends $n$ copies of the same MAC packets in three randomly selected slots, the payload in each burst is encoded by a convolutional \cite{paperCRDSA} or turbo \cite{paperCRDSA++} code of rate $r = \frac{1}{2}$. This is equivalent to a general code of rate $R = \frac{1}{2n}$ with $(n,~k) = (n,~1)$.

The satellite receives on its time frame an interfered signal which is the sum of all modulated bursts sent by the $N_u$ users. In the first round, the receiver seeks the first decodable user. If at least $k$ non-collided (clean) bursts amongst the $n$ ones of an user are identified, the decoding can be realized by the receiver to rebuild the $n$ bursts. Then, the signals corresponding to the $n$ recovered bursts are located and subtracted with PNC processing from the sum. The recovery process of phases and amplitudes is detailed in \cite{analogNC}. After the subtraction, the resulting signal is just the sum of $N_u - 1$ remaining users' signals and the channel noise. The decoding algorithm is iterative until the arrival to a deadlock situation where no user is still decodable.

\begin{figure}
\centering
\includegraphics[width=2.5in]{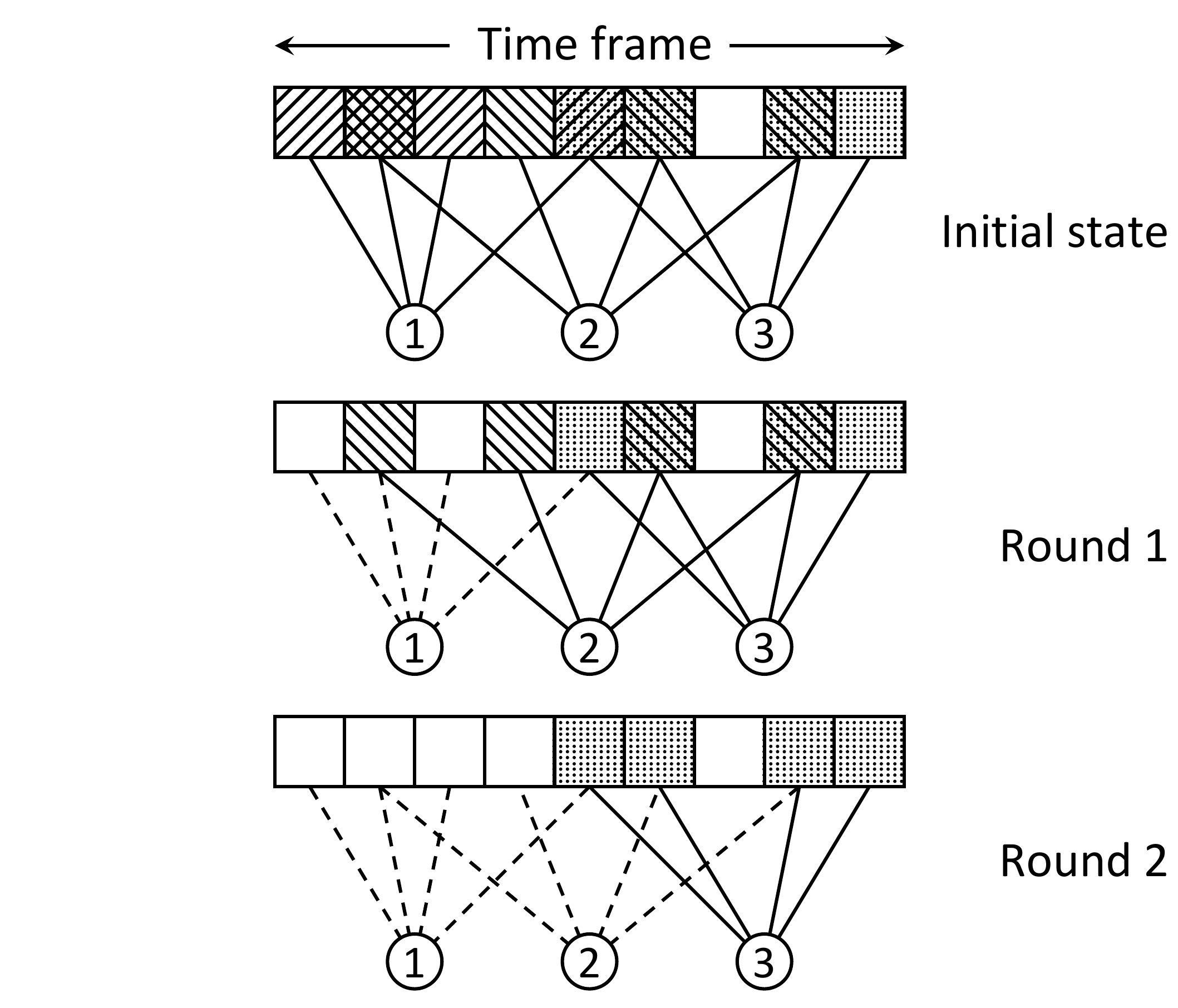}
\caption{Decodable case of decoding algorithm}
\label{exampledecodable}
\end{figure}

\begin{figure}
\centering
\includegraphics[width=2.5in]{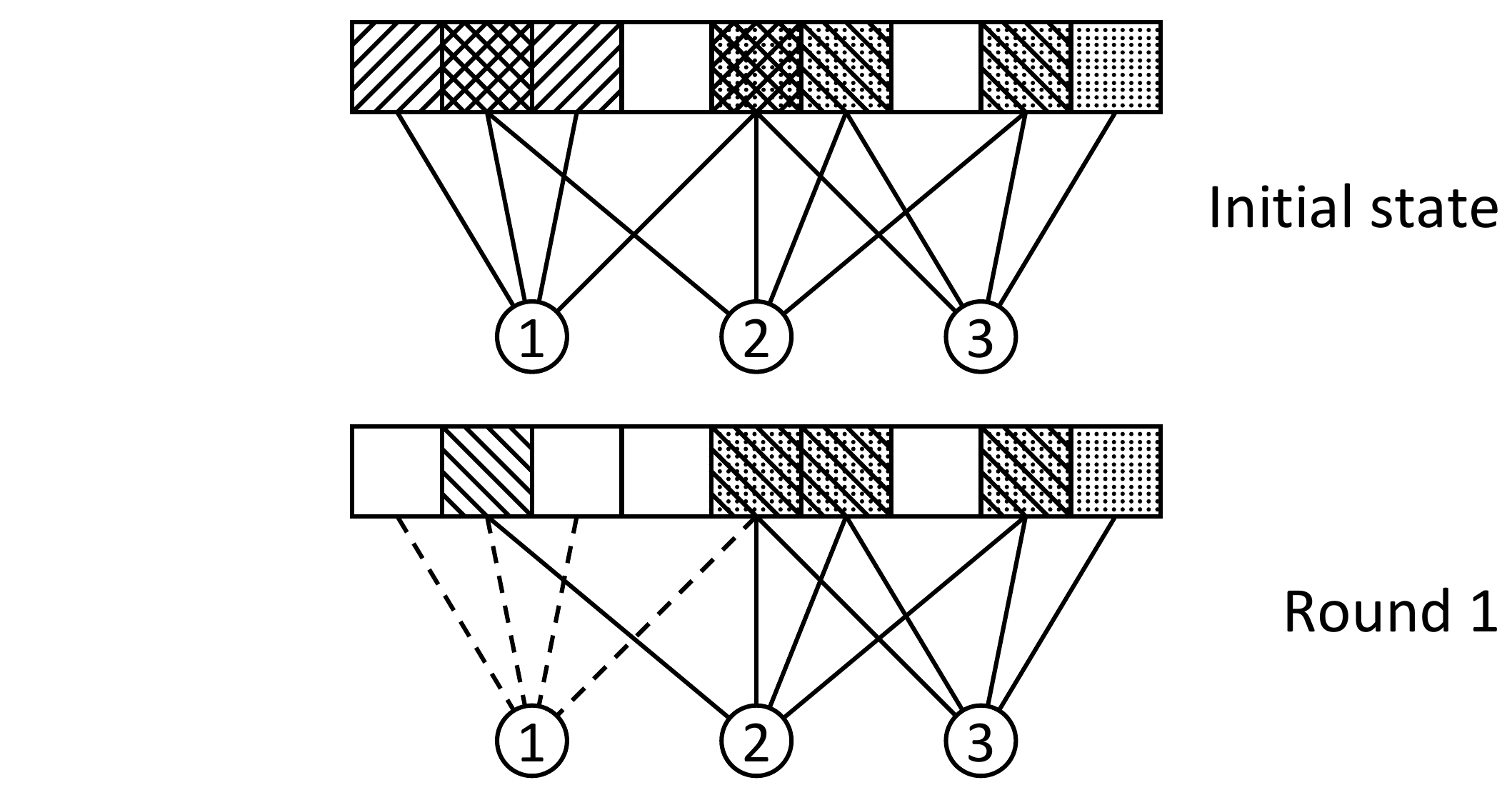}
\caption{Deadlock case of decoding algorithm}
\label{exampleNoNdecodable}
\end{figure}

Figures \ref{exampledecodable} and \ref{exampleNoNdecodable} are two examples of the different rounds of this decoding algorithm. All users have the same parameters $(n,~k) = (4,~2)$. In figure \ref{exampledecodable}, at the reception of the signal, only the data of user 1 can be decoded. After its decoding by the receiver, its four encoded bursts are available and then can be subtracted with PNC operations from the corresponding collided signals (in the $2^{nd}$ and $5^{th}$ time slots). Some bursts of user 2 are no longer in collision and then becomes decodable. On the next round, user 2 is decoded, followed by the user 3 on the 3rd round. Figure \ref{exampleNoNdecodable} shows a case of deadlock between users 2 and 3. It is therefore impossible to decode messages from this frame for both users.
\section{Performance Evaluation}
\label{sec:evaluation}
\subsection{Theoretical Analysis}
\label{sec:TheoreticalAnalysis}
This section focuses on a density evolution approach of the system previously  presented. It can be observed that the principle of the decoding algorithm is similar to the one used for LDPC codes on a Binary Erasure Channel (BEC) \cite{ldpccode:introduction}. These algorithms are iterative and called message passing algorithms. In our system, occupied slots and users play respectively the roles of message nodes $m$ and check nodes $c$. At each round of the algorithm, messages are passed from $m$ to $c$ and vice versa in the same way that the messages exchanged in the iterative decoding algorithm of LDPC codes over BEC. A message node is called of degree $d$ if connected to $d$ edges (the corresponding slot contains $d$ bursts). To push forward the analogy with LDPC erasure codes, a message node is considered as erased if $d > 1$. Moreover, a deadlock situation like in Figure \ref{exampleNoNdecodable} is called a stopping set in the LDPC context.

It is well known that a density evolution analysis requires the independence assumption between the decoding of the nodes \cite{ldpccode:introduction}. Although, in our context, this hypothesis is not really verified because the frame size $N_s$ is limited, we apply the density evolution method to our study, and afterward, analyze the gap between the theory and simulations. In our case, a check node $c$ can recover the value of the $n$ bursts if the number of erasures is less than $e = n-k$ even though $e > 1$ as in \cite{gldpc}. Following the assumptions of the access method, the maximum message node degree denoted $d_s$ is equal to the number of users $N_u$ in the system.

To describe the density evolution analysis, we denote by $\alpha_{d,l}$ the fraction of message nodes of degree $d$ at round $l$. We consider in this study that the users may have different couples $(n,~k)$. We denote by $(n_i,~k_i)$ the parameters of the user $i$, for $i=1,\ldots, N_u$. It is not required to calculate the fraction of edges connected to check nodes because the exact distribution of these parameters is known.

We also note:
\begin{itemize}
	\item $P_l$, the probability of a slot/message node to send an erasure at round $l$;
	\item $Q_l$, the probability of a user/check node to send an erasure (the probability of non decoding of a user) at round $l$.
\end{itemize}

In such a system, $P_0$ is simply the initial erasure probability and is computed as follows:
\begin{equation} \label{P0}
P_0 = \frac{\sum_{d = 2}^{d_s} d \alpha_{d,0} N_s}{\sum_{i = 1}^{N_u}n_i}
\end{equation}
At round $l$, the bursts sent by a user are erased with probability $P_l$. Consequently, the probability of decoding of user $j$ using a couple $(n_j,~k_j)-$code  at the same round, denoted $\bar{Q}_{l,j}$ is:
\begin{equation} \label{Qbarrelm}
\bar{Q}_{l,j} = \sum_{i = k_j}^{n_j} {n \choose i} (1-P_l)^i{P_l}^{n_j -i}
\end{equation}
Thus, probability of non decoding in the system at round $l$ may be computed by averaging over all possible values of couples $(n_j, k_j)$, we get:
\begin{equation} \label{Ql}
{Q}_{l} = 1 - \frac{\sum_{j = 1}^{N_u}\sum_{i = k_j}^{n_j} {n \choose i} (1-P_l)^i{P_l}^{n_j -i}}{N_u}
\end{equation}
On the other side, a message node of degree $d$, connected to check nodes $c$, $c_1$, ..., $c_{d-1}$ sends an erased message to $c$ if it received erasures from one of others check nodes $c_1$, ..., $c_{d-1}$ on the previous round. Thus, the probability of a message to be erased $P_{l+1}$ at round $l+1$ is:
\begin{equation} \label{Plplus1}
P_{l+1} = \frac{\sum_{d = 2}^{d_s} d \alpha_{d,l} N_s}{(\sum_{i = 1}^{N_u}n_i) (1 - \frac{l}{N_u})} \beta_l + P_l (1 - \beta_l)
\end{equation}
where $\beta_l$ is the probability that a user is decoded at the previous round, computed with respect to $Q_i$, where $i$ varies from 1 to $l$.

Hence, the erasure probability $P_l$ can be calculated for each round $l\geq0$ by using (\ref{P0}), (\ref{Qbarrelm}), (\ref{Ql}) and  (\ref{Plplus1}). If $\lim\limits_{l \to N_u} P_l = 0$, the packet loss ratio ($PLR$) at the end of the algorithm is therefore zero.

\begin{figure}
\centering
\includegraphics[width=3.3in]{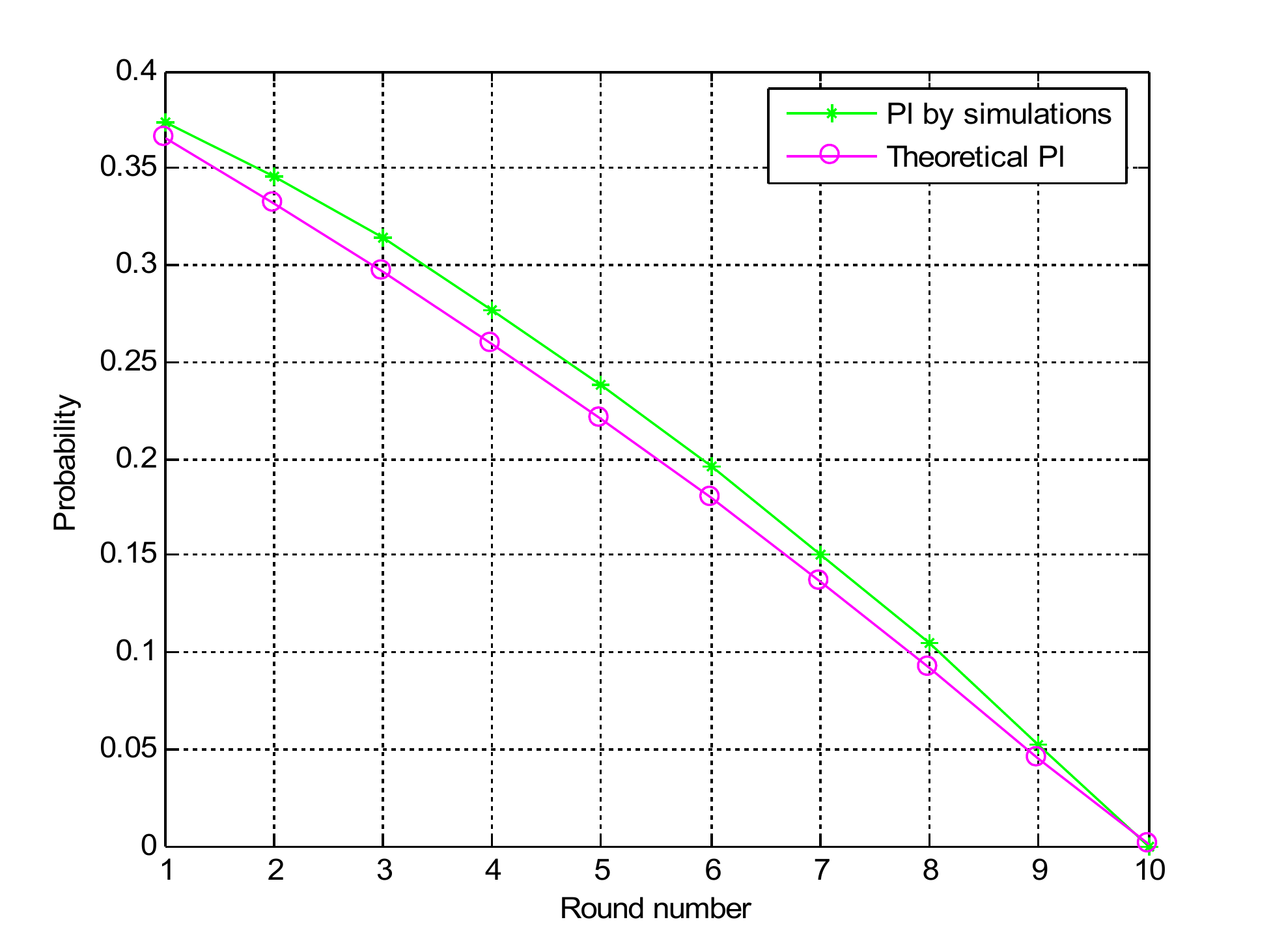}
\caption{Probability of a burst to be erased, RS code (5, 2), frame size = 100, 10 users}
\label{Pl_10_5_2_100}
\end{figure}

\begin{figure}
\centering
\includegraphics[width=3.3in]{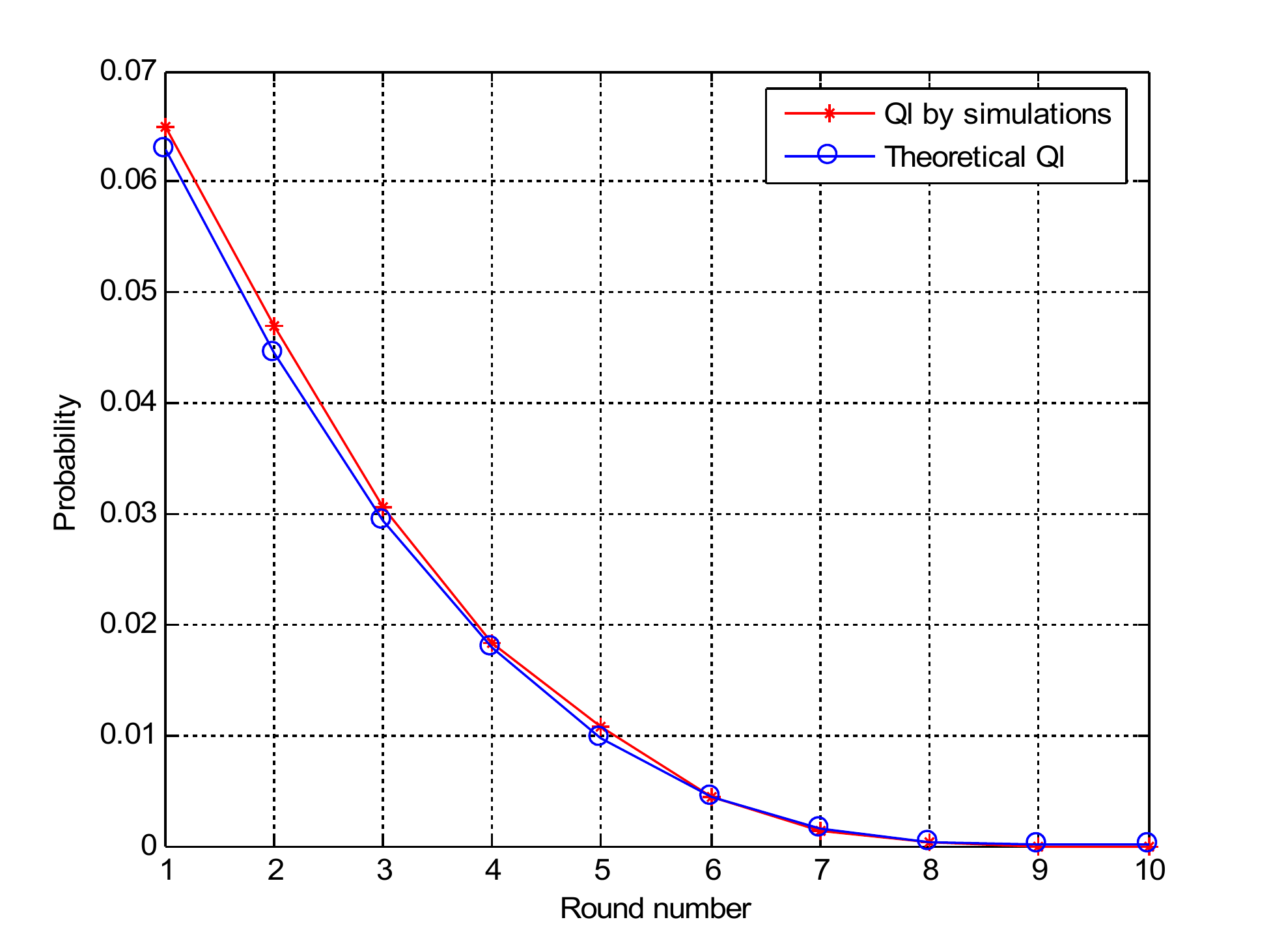}
\caption{Probability of a user node to send an erasure, RS code (5, 2), frame size = 100, 10 users}
\label{Ql_10_5_2_100}
\end{figure}

The validation of theoretical formulas above has been confirmed by comparison with simulation results. Figures \ref{Pl_10_5_2_100} and \ref{Ql_10_5_2_100} give the values of $P_l$ and $Q_l$ at each round of the decoding algorithm for a system with 10 users using the same error-correcting code and $(n,~k) = (5,~2)$ and with a frame size of 100 slots. In this case, 50 bursts are sent. This means that less than a half of slots of the frame are occupied. The theoretical curves and those obtained by simulations are very close.
In a system with 20 users, 100 slots and $(n,~k) = (4,~2)$ used by all users, the number of bursts sent is relatively large compared to the length of the frame. This configuration strongly contradicts the independence assumption, and partly explains a gap between theory and simulations.


Following studies investigate "extreme" conditions where the analysis of density evolution are less valid. We seek to fill up the frame without saturating the slots in order to find the best decoding rate of the system. Therefore, only results from simulations will be detailed and presented.
%
%
%
%
\subsection{Simulation Results}
\label{sec:TheoreticalAnalysis}
In the following study, we use the concepts  of normalized throughput and normalized load used in \cite{paperCRDSA}. The normalized throughput is denoted by $T$. Its computation is based on the total number of $k_i$ of all users and the final Packet Loss Ratio ($PLR$). We get:
\begin{equation} \label{normalizedThroughput}
T = \frac{\sum_{i = 1}^{N_u}k_i(1 - PLR)}{N_s}
\end{equation}
We denoted by $G$ the normalized load which is defined as the mean number of bursts that is necessary to decode transmitted in a slot. Thus, 
\begin{equation} \label{normalizedLoad}
G = \frac{\sum_{i = 1}^{N_u}k_i}{N_s}
\end{equation}
It is important to recall that the noise on the transmission channel is considered as to be absorbed by the error-correcting code and the non-decoding is only due to collisions. 

The value of $G$ that maximizes $T$ must be chosen carefully. Indeed, $T$ is bounded by $G$ so with a greater $G$, the expected normalized throughput can be higher. But if $G$ exceeds a certain value, the frame is full, the collision rate is high, which causes a large packet loss ratio and reduced $T$.

%
%
%
\begin{figure}
\centering
\includegraphics[width=3.3in]{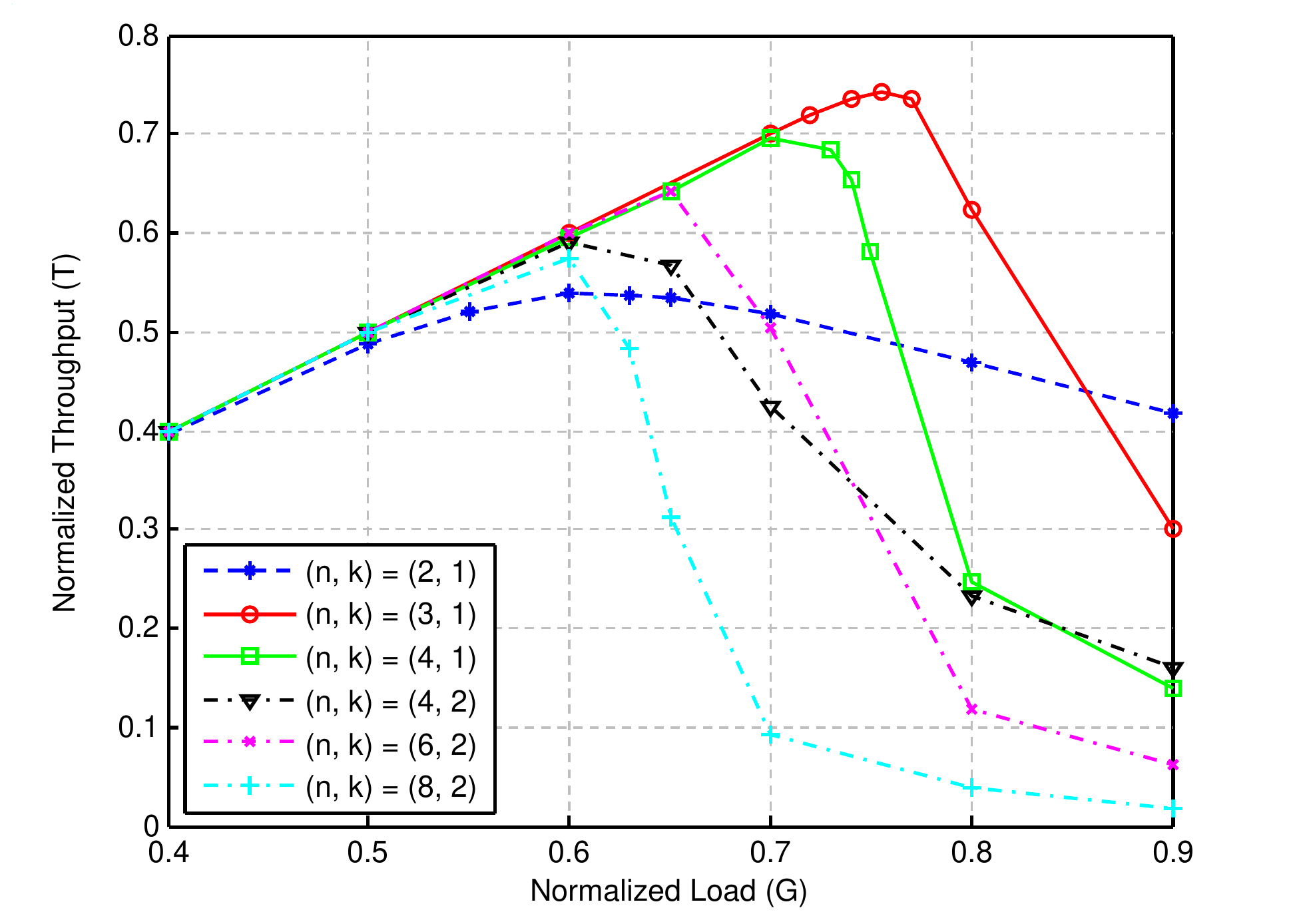}
\caption{Normalized throughput versus normalized load, $N_s$ = 400}
\label{variationG}
\end{figure}
The Figure \ref{variationG} gives examples of $T$ versus $G$ for a frame size equal to 400. We can also observe in this figure the maximum normalized throughput (for this frame size) of different set of parameters $n,~k)$. The optimal throughput of 0.7433 for a normalized load = 0.755 is offered by the couple $(n,~k) = (3,~1)$. These results are conformed to the ones of analysis realized on CRDSA* \cite{invitedpaperCRDSA++}.

\begin{figure}
\centering
\includegraphics[width=3.3in]{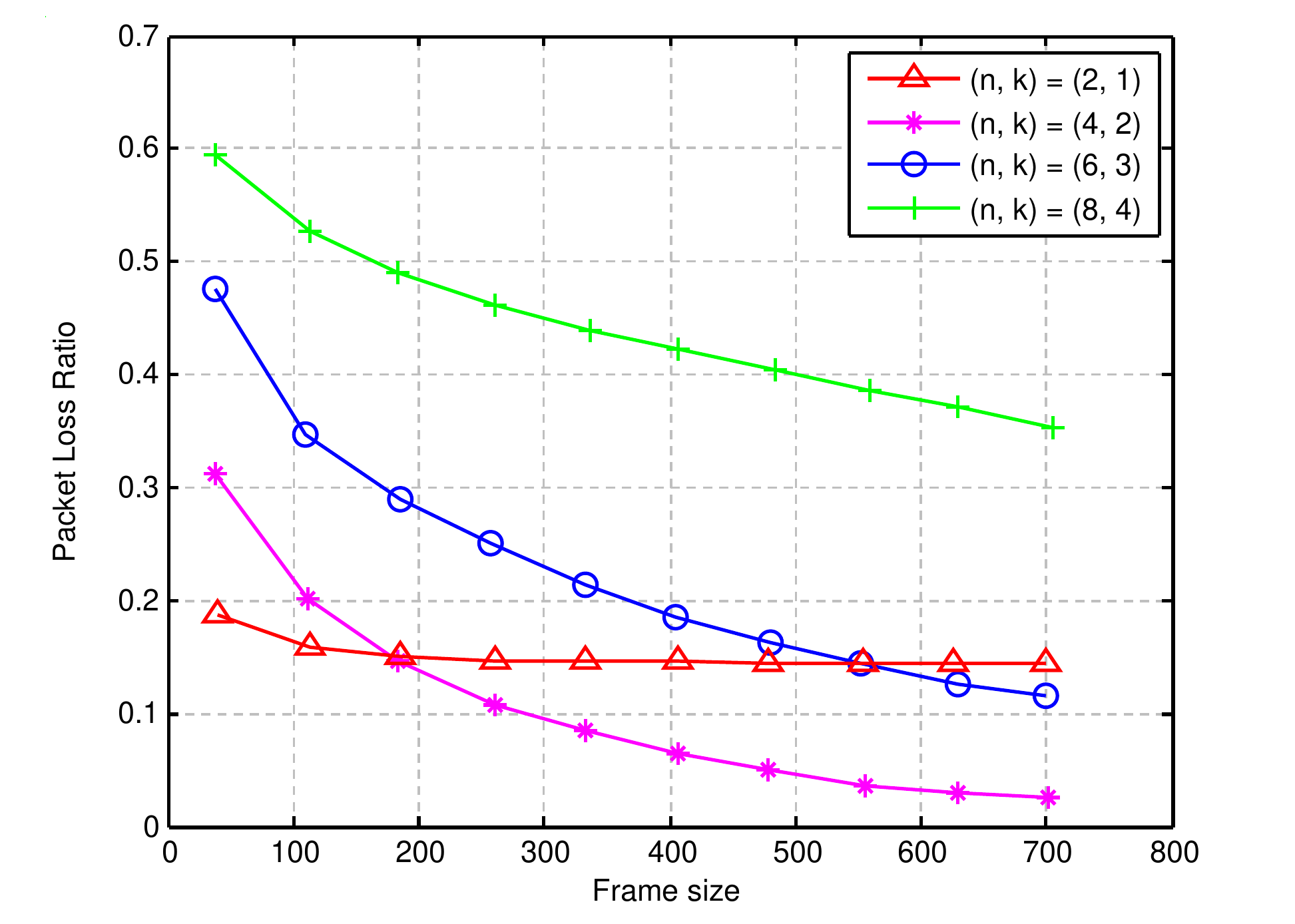}
\caption{Packet Loss Ratio for Normalized Load = 0.63, $N_s$ = 700}
\label{PLR_063_2_1}
\end{figure}

\begin{figure}
\centering
\includegraphics[width=3.3in]{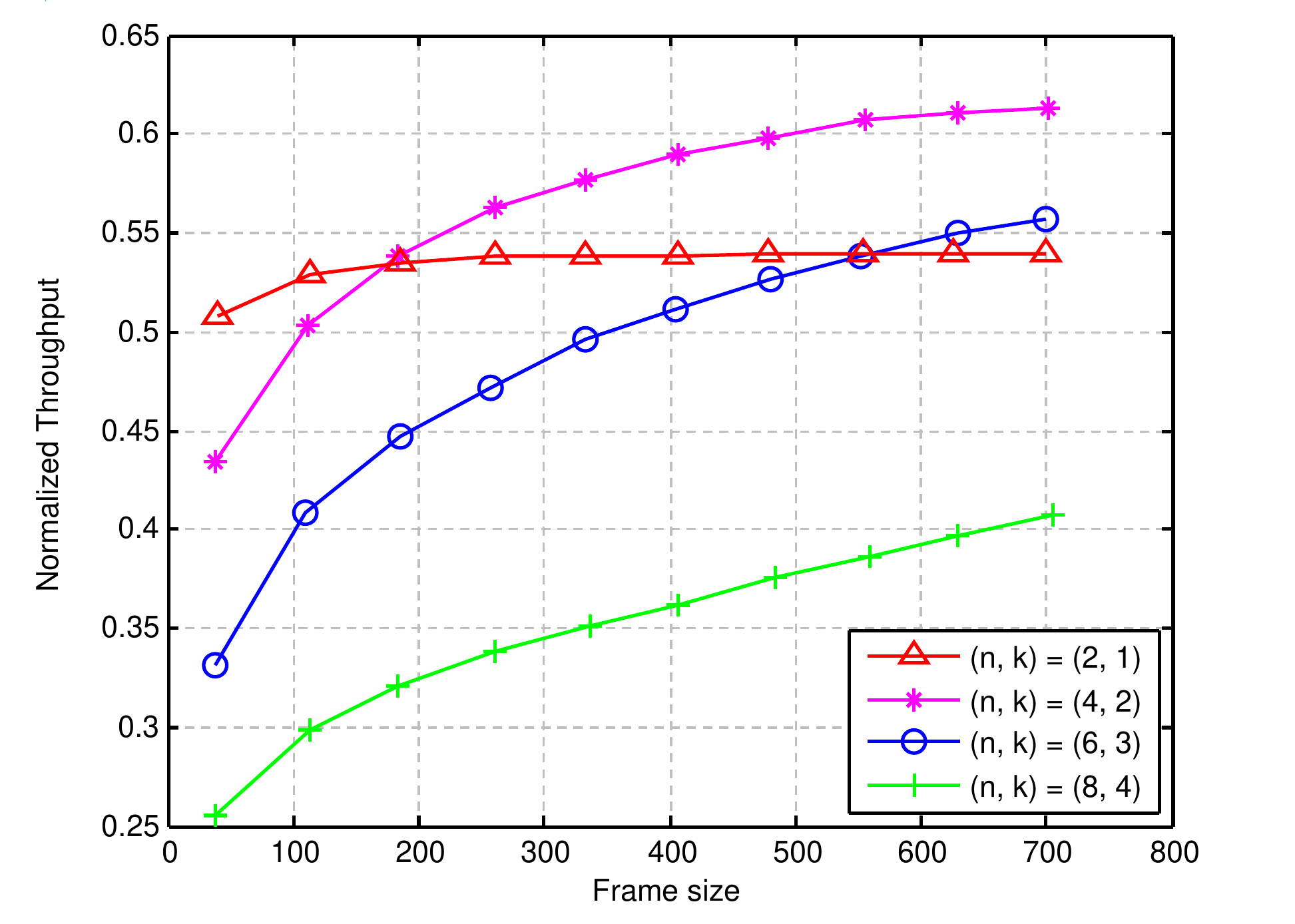}
\caption{Normalized Throughput for Normalized Load = 0.63, $N_s$ = 700}
\label{NormalizedThroughput_063_2_1}
\end{figure}

In Figure \ref{NormalizedThroughput_063_2_1}, we evaluate several sets of parameters $(n,~k)$ proportional to $1/2$. For each set, we fix the normalized load $G$, and we vary the frame size to find parameters that maximize $T$. Of course, since $G$ is fixed and $N_s$ varies, the number of users in the system $N_u$ varies and is computed as  $N_sG/k$. We can observe that the system using the couple $(4,~2)$ approaches a maximum normalized throughput of 0.6155 for a normalized load $G = 0.63$ and the system offers a $PLR$ of 0.0264 (see Figure \ref{PLR_063_2_1}).

Figures \ref{PLR_063_2_1} and \ref{NormalizedThroughput_063_2_1} show that for the same normalized load and the same parameters $(n,~k)$, a lower packet loss ratio is achieved when the frame size and number of users increases ($N_u$ is proportional to $N_s$).  
\section{Discussion}
\label{sec:discussion}
The results presented in the last section seem promising because the obtained normalized throughput significantly outperforms the one obtained by classical slotted-ALOHA ($0.7433$ vs. $0.37$). In the optimum case, the parameters correspond to those analyzed in the CDRSA system. It validates once again a part of our study. However, we investigate a possibility of not using a repetition code, but a more sophisticated code that gives more flexibility in the choice of the parameters. With the CRDSA protocol, it is mandatory to use $k = 1$. Combinations such as $(3,~2)$, $(6,~3)$ are not possible. These ones should be more efficient because for the same ratio, a long code is more efficient than a short one.


Moreover, up to now, the couples $(n,~k)$ used are the same for all users in the communication system. This condition is not necessary and a configuration with several code rates and parameters may provide better results in terms of global normalized throughput. Indeed, according to LDPC codes theory \cite{ldpccode:introduction}, the best decoding performance are obtained for irregular codes, i.e. for codes with particular distributions of degrees on the left and right nodes. Following this idea, studies on density evolution can be used to define the distributions that optimize our decoding.

Another future important study will concern the use of collided bursts in the decoding process. In the present paper, these burst are considered as erased, however from a information theoretic point of view, these bursts, although submitted to a strong level of interference, can significantly improve the decoding performance.  
\section{Conclusion}
\label{sec:conclusion} 
In this paper, we have presented a random multiple access strategy designed for a slotted-ALOHA channel. The main originality of our scheme is to combine diversity transmission of coded data bursts and successive interference cancellation technique. This can be seen as a generalization of CRDSA*. We showed that the behavior of the decoding process can be modeled with a density evolution analysis. Simulation results showed a better throughput achieved over a satellite link compared to standard slotted ALOHA. Several promising extensions of these first results are planned in order to improve the performance of the proposed scheme.
\section*{Acknowledgments}
The authors would like to thank Raoul Pr\'evost and Emmanuel Lochin for their help.

\bibliographystyle{IEEEtran}
\bibliography{biblio}

\end{document}